\documentclass[preprint,amsmath,amssymb]{revtex4}

\usepackage{graphics}
\usepackage[dvips]{color}
\usepackage{amsthm}

\newtheorem{lemm}{Lemma}
\newtheorem{pro}{Proof}
\newtheorem{theo}[lemm]{Theorem}

\newtheorem{coll}[lemm]{Corollary}

\begin{document}

\title{On the size-consistency of the reduced-density-matrix method and the unitary invariant diagonal $N$-representability conditions}

\author{Maho Nakata}
\email{maho@riken.jp}
\affiliation{Advanced Center for Computing and Communication, RIKEN, 2-1 Hirosawa, Wako-city, Saitama, 351-0198, Japan}

\begin{abstract}
Variational calculation of the ground state energy and its properties using the second-order reduced density matrix (2-RDM) is a promising approach for quantum chemistry. A major obstacle with this approach is that the $N$-representability conditions are too difficult in general. Therefore, we usually employ some approximations such as the $P$, $Q$, $G$, $T1$ and $T2^\prime$ conditions, for realistic calculations. The results of using these approximations  and conditions in 2-RDM are comparable to those of CCSD(T). However, these conditions do not incorporate an important property; size-consistency. Size-consistency requires that energies $E(A)$, $E(B)$ and $E(A\cdots B)$ for two infinitely separated systems $A$, $B$, and their respective combined system $A\cdots B$, to satisfy $E(A\cdots B) = E(A) + E(B)$.
In this study, we show that the size-consistency can be satisfied if 2-RDM satisfies the following conditions: (i) 2-RDM is unitary invariant diagonal $N$-representable; (ii) 2-RDM corresponding to each subsystem is the eigenstate of the number of corresponding electrons; and (iii) 2-RDM satisfies at least one of the $P$, $Q$, $G$, $T1$ and $T2^\prime$ conditions.
\end{abstract}

\maketitle

\section{Introduction}
We are interested in direct determination of the second-order reduced density matrix (2-RDM) as an alternative to using the wavefunction, because 2-RDM contains all the relevant information for the ground state energy and physical properties of the $N$-particle (electronic) system in a compact form \cite{DM1960}. In 2-RDM formulation, the ground state energy is obtained by minimizing the Hamiltonian, which is a linear functional of 2-RDM. We call such an approach as ``the RDM method''. 

The $N$-representability conditions in 2-RDM \cite{Coleman63} are the key to the RDM method and essential for electron correlation. However, complete establishment of the 2-RDM seems impractical to date \cite{Garrod64,Liu07}. As such, we usually employ and investigate some of the necessary conditions such as the $P$, $Q$ \cite{Coleman63}, $G$ \cite{Garrod64}, $T1$, and $T2$ \cite{Zhao04,Erdahl78} conditions. All these linearly positive semidefinite conditions involve the 2-RDM; Erdahl and Jin have previously extended these representability conditions that involve the higher-order RDMs \cite{ERDAHLJIN00}.

The first calculation employing the $P$, $Q$ and $G$ conditions as approximate $N$-representability conditions was done by Garrod {\it et al.}, and they have reproduced very good ground state energy for the Be atom \cite{Garrod76,Garrod75}. However, when Mihailovi\'c {\it et al.} applied these conditions to nuclear structure calculation \cite{Mihailovic75} with disappointing results, researchers thereafter seemed to lose their motivation.

About 25 years later, in 2001, Nakata {\it et al.} formulated the problem in a primal semidefinite program, and performed direct variational calculation of the 2-RDM employing $P$, $Q$ and $G$ conditions using a well-established semidefinite programming solver called SemiDefinite Programming Algorithm (SDPA) \cite{SDPA} and applied the method to many small atoms and molecules \cite{Nakata01,Nakata02}. Their results with the $P$, $Q$, and $G$ conditions were very encouraging; yielding $100$ to $130\%$ correlation energy, and they even reproduced the dissociation curve of the nitrogen dimer in good agreement with fullCI. When Zhao {\it et al.} subsequently  
reformulated and applied the $T1$ and $T2$ conditions to small molecules \cite{Zhao04}, their results are comparable to CCSD(T) at equilibrium geometries. The calculated correlation energies range typically from 100 to 101\%, and these results are evidently better than using $P$, $Q$ and $G$ conditions \cite{Zhao04,Fukuda07,Nakata08}. Mazziotti immediately confirmed the findings of Nakata {\it et al.} \cite{Nakata01,Nakata02} and Zhao {\it et al.} \cite{Zhao04}, and applied these conditions to larger systems by implementing a semidefinite programming solver \cite{Mazziotti02,MazziottiBook,Mazziotti04}.  An attempt by Mazziotti {\it et al.} in massive application to verify these $N$-representability conditions is in fact important. Later, Braams {\it et al.} and Mazziotti found $T2^\prime$ condition, which slightly improves $T2$ condition \cite{Braams07, MazziottiT2bar}.

Therefore, necessary conditions such as $P$, $Q$, $G$, $T1$ and $T2^\prime$ seemed very promising; however, in 2009, Van  Aggelen {\it et al.} reported a drastic failure of the dissociation limits of some molecules on dissociation \cite{Aggelen09}. As they observed fractionally charged atoms, they become aware of $P$, $Q$ and $G$ conditions lacked size-consistency. For example, at the dissociation limit of $\rm CN^-$, the Mulliken population of $\rm C$ and $\rm N^-$ were $6.60$ and $7.40$, respectively. When Nakata {\it et al.} examined size-consistency of 32 non-interacting methane and nitrogen molecules under $P$, $Q$ and $G$ conditions, non-size-consistent contributions were $3\times 10^{-4}$ and $3\times 10^{-3}$ atomic unit \cite{Nakata09}. Verstichel {\it et al.} proposed a method to fix these pathological behaviors by imposing fractional $N$-representability conditions to the subsystems with numerical verifications, and applied them to the dissociation limit of $\rm BeB^+$, $\rm N_2$ and isoelectronic molecule series \cite{Verstichel10,Aggelen10}. Although useful in some cases, this method would not fix non-size-consistency problems for the all cases. Moreover, calculation of various electron numbers for the subsystems is required, and the choice of subsystems is unclear, especially when a chemical reaction occurs \cite{Nakata09}.

In this study, we show condition adequate for establishing size-consistency; an approximately $N$-representable variational space is size-consistent when 2-RDM satisfies the unitary invariant diagonal $N$-representability, and positive semidefinite type $N$-representability conditions, assuming that the calculated 2-RDM for each subsystem is an eigenstate of the number of electrons. The diagonal $N$-representability conditions of 2-RDM have been extensively studied by Yoseloff, Kuhn, Davidson, McRae and Pistol \cite{DIAGONALREP,AyersDavidson07,MEPistol}. Useful features of these conditions are that they can be written only by linear inequalities of the diagonal elements of 2-RDM, and if a given 2-RDM is diagonal $N$-representable then there exists certain wavefunction that reduces to the diagonal elements of 2-RDM. The expression of these linear equations explicitly depends on which one-particle basis is employed. Thus, even though a trial 2-RDM is diagonal $N$-representable, it is not necessary diagonal $N$-representable on a different basis. Therefore we require its invariance, which can be included via usual semidefinite formulation.

Our result yields a sufficient condition. Aside from significant theoretical importance, it is still impractical as incorporation of the diagonal $N$-representability conditions in the RDM method has not yet been attempted, and the assumption that each subsystem being the eigenstate of the number of electrons is not trivial as it is violated in practice \cite{Aggelen09}. 

This paper is organized as follows. In Section II, we briefly review the RDM method and $N$-representability conditions. The main results are shown in Section III. The discussion and the conclusion are placed in Section IV.

\section{The reduced-density-matrix method and the $N$-representability conditions}
In this section, we briefly describe the RDM method and $N$-representability conditions. Since the Hamiltonian involves one- and two-particle interactions, the Hamiltonian is thus expressed as follows:
\[
H = \sum_{ij} v^i_j a^\dagger_i a_j+\frac{1}{2} \sum_{i_1 i_2 j_1 j_2} w^{i_1 i_2}_{j_1 j_2} a^\dagger_{i_1} a^\dagger_{i_2} a_{j_2} a_{j_1},
\]
where $a^\dagger$ and $a$ are creation and annihilation operators, and $v$ and $w$ are one-particle and two-particle operators, respectively.
The first- (1-RDM; $\gamma$) and second-order reduced density matrix (2-RDM; $\Gamma$) are respectively defined as follows:
\[
 \gamma^{i}_{j} = \langle \Psi | a_i^\dagger a_j | \Psi \rangle,
\]
and
\[
 \Gamma^{i_1 i_2}_{j_1 j_2} = \frac{1}{2}\langle \Psi | a_{i_1}^\dagger a_{i_2}^\dagger a_{j_2} a_{j_1} | \Psi \rangle.
\]
We can alternatively define using an ensemble state; $\sum_{p}w_p |\Psi_p\rangle\langle \Psi_p|$, where $w_p$ is a non-negative value with unit sum, and $\{ |\Psi_p\rangle\} $ is a complete orthonormal set of $N$-particle space to define 1- and 2-RDMs.
The ground state energy $E_g$ can be obtained using the 1- and 2-RDMs as follows:
\begin{eqnarray}
E_g & = & \min_\Psi \langle \Psi | H | \Psi \rangle \label{wf} \\
    & = & \min_\Psi \left \{ \sum_{ij} \langle \Psi | v^i_j a^\dagger_i a_j | \Psi \rangle +\frac{1}{2} \sum_{i_1 i_2 j_1 j_2} \langle \Psi | w^{i_1 i_2}_{j_1 j_2} a^\dagger_{i_1} a^\dagger_{i_2} a_{j_2} a_{j_1} | \Psi \rangle \right \} \nonumber \\
    & = & \min_{\Gamma, \gamma} \left \{ \sum_{ij} v^i_j \gamma^i_j + \sum_{i_1 i_2 j_1 j_2} w^{i_1 i_2}_{j_1 j_2} \Gamma^{i_1 i_2}_{j_1 j_2} \right \} \label{rdm}.
\end{eqnarray}

In this way, $E_g$ be calculated by 1- and 2-RDMs. We refer such an approach as ``the RDM method''. 

Until the 1960s, researchers tried to minimize the 2-RDMs using eqn. (\ref{rdm}) instead of eqn. (\ref{wf}). However, energy values obtained are far lower than the actual values, because their trial 2-RDMs are not derived from wavefunctions. 
The condition for 2-RDM to actually derived from a wavefunction is called 
the ``$N$-representability condition'', which was first formulated by Coleman \cite{Coleman63}. This condition states that if a given arbitrary 2-RDM satisfies this condition, and then it should have been derived from a wavefunction, but not otherwise. If it is non-$N$-representable, no wavefunction is reduced to the 2-RDM. 

We denote ${\cal E}_N$ as the $N$-representable set of the 1- and 2-RDM, and the RDM method is then expressed as follows:
\begin{eqnarray}
E_g & = & \min_{{\cal E}_N \ni \Gamma, \gamma} \left \{ \sum_{ij} v^i_j \gamma^i_j + \sum_{i_1 i_2 j_1 j_2} w^{i_1 i_2}_{j_1 j_2} \Gamma^{i_1 i_2}_{j_1 j_2} \right \}.
\end{eqnarray}

Here are some of the trivial $N$-representability conditions:
\begin{description}
\item {(a)} 1-RDM and 2-RDM are Hermitian,
\[
\gamma^i_j = ( \gamma^j_i )^* , \quad \Gamma^{i_1 i_2}_{j_1 j_2} = ( \Gamma^{j_2 j_1}_{i_1 i_2} )^*,
\]

\item {(b)} 2-RDM is antisymmetric,
\[
\Gamma^{i_1 i_2}_{j_1 j_2} = - \Gamma^{i_2 i_1}_{j_1 j_2} = -\Gamma^{i_1 i_2}_{j_2 j_1} = \Gamma^{i_2 i_1}_{j_2 j_1},
\]

\item {(c)} the trace conditions are valid, or equivalently, the system is an eigenstate of the number of the electrons,
\[
\sum_{i=1} \gamma^i_i = N, \quad \sum_{i,j=1}\Gamma^{ij}_{ij} = \frac{N(N-1)}{2},
\]

\item {(d)} a partial trace condition exists between 1- and 2-RDM,
\[
\frac{N-1}{2}\gamma^i_j = \sum_{k=1}\Gamma^{ik}_{jk}.
\]
\end{description}
From (c), we can prove $\langle N^2 \rangle - \langle N \rangle ^2=0$. In other words, the system is an eigenstate of the number of electrons. Of course, the above list is not exhaustive. For 1-RDM, the complete $N$-representability conditions are quite easily obtained; a 1-RDM is $N$-representable if it satisfies the conditions stated above, and its eigenvalues are
between zero and one \cite{Kuhn60,Coleman63}. However, in the case of 2-RDM, obtaining a complete set of conditions is impractical \cite{Garrod64, Liu07}. Thus, finding a physically meaningful approximation or interpretation of the $N$-representability conditions is crucial for the RDM methods. Some well-known necessary conditions are the $P$, $Q$ \cite{Coleman63}, and $G$ \cite{Garrod64} defined by the positive semidefiniteness of $P$, $Q$, and $G$-matrix which in turn are defined as follows:
\begin{eqnarray}
P^{i_1 i_2}_{j_1 j_2} & = &\langle \Psi | a_{i_1}^\dagger a_{i_2}^\dagger a_{j_2}a_{j_1}    | \Psi \rangle \succeq 0 \nonumber \\
Q^{i_1 i_2}_{j_1 j_2} & = &\langle \Psi | a_{i_1} a_{i_2} a_{j_2}^\dagger a_{j_1}^\dagger   | \Psi \rangle \succeq 0, \nonumber \\
G^{i_1 i_2}_{j_1 j_2} & = &\langle \Psi | a_{i_1}^\dagger a_{i_2} a_{j_2}^\dagger a_{j_1}   | \Psi \rangle \succeq 0, \nonumber
\end{eqnarray}
where $A \succeq 0$ denotes $A$ is a positive semidefinite matrix; i.e. the eigenvalues of $A$ are non-negative. For a system $A\cdots B$, we denote ${\cal \tilde E}_N^{A\cdots B}$ as 
\[
{\cal \tilde E}_N^{A\cdots B}= \{(\Gamma,\gamma)| \mbox{$\Gamma$ satisfies the $P$, $Q$, and $G$ conditions with trivial $N$-rep.conditions from (a) to (d). } \}
\]
Using the $P$, $Q$, and $G$ conditions, the RDM method can be formulated as a positive semidefinite programming problem \cite{Nakata01,SDPA}. Then, the RDM method is represented as: 
\begin{eqnarray}
E_g & = & \min_{{\cal \tilde E}_N^{A\cdots B} \ni \Gamma, \gamma} \left \{ \sum_{ij} v^i_j \gamma^i_j + \sum_{i_1 i_2 j_1 j_2} w^{i_1 i_2}_{j_1 j_2} \Gamma^{i_1 i_2}_{j_1 j_2} \right \}.
\end{eqnarray}

In 2004, Zhao {\it et al.} \cite{Zhao04} implemented additional $N$-representability conditions called the $T1$ and $T2$ conditions. Later, Braams {\it et al.} \cite{Braams07} and Mazziotti \cite{MazziottiT2bar} derived $T2^\prime$ condition, which replaces $T2$ with slight enhancement.
These conditions state that $T1$, $T2$ and $T2^\prime$ matrices are positive semidefinite, and can also be written as a linear functional of 1- and 2-RDMs. Therefore, we can still formulate as a standard type of semidefinite programming problem. In the RDM method, we usually obtain non-physical 2-RDM, in terms of having no wavefunctions that can be reduced to 2-RDM. Nevertheless, deviations seem to be negligible for the equilibrium geometry of molecules \cite{Nakata01,Mazziotti02,Zhao04,Mazziotti04,Fukuda07,MazziottiBook,Nakata08}. 

Other well-known $N$-representability conditions include the diagonal $N$-representability conditions 
\cite{DIAGONALREP,AyersDavidson07}.
These conditions state that if a trial 2-RDM $\Gamma$ satisfies the diagonal $N$-representability conditions, then there exist certain wavefunctions $|\Psi\rangle$ which can be reduced to the diagonal part of original 2-RDM. 
\begin{equation}
\Gamma^{ij}_{ij} =  \frac{1}{2} \langle \Psi | a^\dagger_i a^\dagger_j a_j a_i | \Psi \rangle  \label{diagnrep}
\end{equation}
Note that such a 2-RDM may have its ancestor wavefunctions $|\Phi\rangle$
different from $|\Psi\rangle$. Explicitly, we can construct $|\Phi\rangle$ as follows:
\begin{equation}
| \Phi \rangle = \sum_i \sqrt{w_i} e^{i\phi_i} | \Phi_i\rangle, \label{purepreimag}
\end{equation}
where $ | \Phi_i \rangle$ are the Slater determinants, $\phi_i$ is an arbitrary phase factor and $w_i$ are non-negative numbers with unit sum ($\sum_i w_i = 1, w_i \geq 0 $) \cite{AyersDavidson07}. In this case, non-diagonal elements by $|\Phi\rangle$ and $|\Psi\rangle$ are likely to be different.

Another important feature of the diagonal $N$-representability conditions is that they can be written as a linear inequality of the diagonal elements of 2-RDM. Here we list some inequalities:
\begin{eqnarray}
&& \gamma^i_i \geq 0 \nonumber \\
&& 1 - \gamma^i_i \geq 0 \nonumber \\
&& \Gamma^{ij}_{ij} \geq 0 \nonumber \\
&& \gamma^i_{i}  - \Gamma^{ij}_{ij} \geq 0 \nonumber \\
&& 1 - \gamma^i_i - \gamma^j_j  + \Gamma^{ij}_{ij} \geq 0. \nonumber
\end{eqnarray}
Although all the inequalities can be written, expressing the complete set of diagonal representability conditions may be impractical, as the number of inequalities grow astronomically with size of the basis and the number of electrons \cite{DIAGONALREP}. In this paper, we do not add any new diagonal $N$-representability condition, and we invesigate this physical nature.
 
The expression of all these inequalities depends on which one-particle basis is employed. Thus, if a non $N$-representable with a diagonal $N$-representable 2-RDM is given,
$\gamma^i_i \geq 0$ is satisfied for the specific choice of one-particle basis. In such a 2-RDM with a different one-particle basis representation, $\gamma^i_i\geq 0$ may be violated. Derivation of inequalities by the diagonal $N$-representability conditions is independent of the choice of the one-particle basis. Therefore, we can generate more $N$-representability conditions by applying a unitary rotation to the one-particle basis. In so doing, if the diagonal $N$-representability conditions with unitary invariant are established \cite{AyersDavidson07}; e.g. if $\gamma^i_i \geq 0$ and $1 - \gamma^i_i \geq 0$ are unitary invariants, these two inequalities impose a constraint on eigenvalues of $\gamma$, which should be in $[0, 1]$. We can prove this as follows: since $\gamma$ is Hermitian, we can diagonalize $\gamma$. On the basis of $\gamma^i_i \geq 0$ $\gamma$ is diagonalized, and the eigenvalues of $\gamma$ are greater or equal to zero. Next, when $1 - \gamma^i_i \geq 0$ generates eigenvalues is smaller or equal to 1, the eigenvalue of $\gamma$ should be within $[0, 1]$ on combining these two results. Interestingly, this is a complete $N$-representability condition of 1-RDM.

Apparently, every linear inequality has its unitary invariant form. If a given 2-RDM is diagonal $N$-representable regardless of the one-particle basis employed, then we call such a 2-RDM as ``unitary invariant diagonal $N$-representable'', and the conditions there of as ``unitary invariant diagonal $N$-representability conditions''.

\section{On the size-consistency}
In this section, after reviewing (i) the definitions of size-consistency and subsystem 1-, and 2-RDM and (ii) lemma and theorem for matrices, we obtain the main result to provide a proof (iii) using these lemma and theorem, if a variational set of 2-RDM satisfies the unitary invariant diagonal $N$-representability conditions and positive semidefinite conditions such as the $P$, $Q$, $G$, $T1$ and $T2^\prime$. The 2-RDM of each subsystem derived is the eigenstate of the number of electrons, and the RDM method with this variational set is then size-consistent. Note that start formulating from using 1-RDM and cumulant of 2-RDM \cite{cumulants}, which are are additively separable quantities, do not solve the problem. Since considering $N$-representability condition on 2-RDM is same as using these quantities. As they contain exactly same information.

\subsection{The definition of size-consistency for the RDM method and subsystems}

Size-consistency may be defined as follows: assuming that $A$ and $B$ are two non-interacting systems. If a variational space is size-consistent, the total energy $E(A \cdots B)$ of the total system $A \cdots B$ is expressed as follows:
\begin{equation}
E(A \cdots B) = E (A) + E(B),
\end{equation}
where $E(A)$ and $E(B)$ are energy values of the subsystems calculated separately \cite{Szabo}.

Note that, as is already known, a variational method using 2-RDM with the $P$, $Q$ and $G$ condition as $N$-representability conditions is not size-consistent \cite{Aggelen09,Nakata09,Aggelen10,Verstichel10}.

First, let us consider the subsystems: 1- and 2-RDM of subsystems $A$ ($\gamma^A, \Gamma^A$) and $B$ ($\gamma^B$
 and $\Gamma^B$) are defined using a set of one-particle basis allocated in each subsystem $A$ or $B$ from the whole system as follows:
\[
 {\gamma^A}^{i}_{j} := \gamma^{i}_{j} \mbox{ \hspace{1cm} where $i, j$ correspond to the one-particle basis of $A$}
\]
\[
 {\Gamma^A}^{i_1 i_2}_{j_1 j_2} := \Gamma^{i_1 i_2}_{j_1 j_2} \mbox{ \hspace{1cm} where $i_1, i_2, j_1, j_2$ correspond to the one-particle basis of $A$}
\]
\[
 {\gamma^B}^{i}_{j} := \gamma^{i}_{j} \mbox{ \hspace{1cm} where $i, j$ correspond to the one-particle basis of $B$}
\]
\[
 {\Gamma^B}^{i_1 i_2}_{j_1 j_2} := \Gamma^{i_1 i_2}_{j_1 j_2} \mbox{ \hspace{1cm} where $i_1, i_2, j_1, j_2$ correspond to the one-particle basis of $B$}.
\]
For other cases, ${\gamma^A}^{i}_{j} = 0$ is designated for either $i$ or $j$ corresponding to the one-particle basis of $B$, etc. 

We can define subsystem Hamiltonians $H^A$ and $H^B$ as well. First, one-particle operators and two-particle operators of subsystems $A$ and $B$ are expressed as follows:
\[
 {v^A}^{i}_{j} := v^{i}_{j} \mbox{ \hspace{1cm} where $i, j$ correspond to the one-particle basis of $A$}
\]
\[
 {w^A}^{i_1 i_2}_{j_1 j_2} := w^{i_1 i_2}_{j_1 j_2} \mbox{ \hspace{1cm} where $i_1, i_2, j_1, j_2$ correspond to one-particle basis of $A$}
\]
\[
 {v^B}^{i}_{j} := v^{i}_{j} \mbox{ \hspace{1cm} where $i, j$ correspond to the one-particle basis of $B$}
\]
\[
 {w^B}^{i_1 i_2}_{j_1 j_2} := w^{i_1 i_2}_{j_1 j_2} \mbox{ \hspace{1cm} where $i_1, i_2, j_1, j_2$ correspond to the one-particle basis of $B$}.
\]
For other cases, ${w^B}^{i_1 i_2}_{j_1 j_2} = 0$ is designated for convenience; if one of $i_1$, $i_2$, $j_1$ or $j_2$ corresponds to one-particle basis of $A$, etc. 
$H^A$ and $H^B$ are then express as follows:
\[
 H^A = \sum_{ij} {v^A}^{i}_{j} a^\dagger_i a_j + \sum_{i_1 i_2 j_1 j_2} {w^A}^{i_1 i_2}_{j_1 j_2}a^\dagger_{i_1} a^\dagger_{i_2} a_{j_2} a_{j_1}
\]
and 
\[
 H^B = \sum_{ij} {v^B}^{i}_{j} a^\dagger_i a_j + \sum_{i_1 i_2 j_1 j_2} {w^B}^{i_1 i_2}_{j_1 j_2}a^\dagger_{i_1} a^\dagger_{i_2} a_{j_2} a_{j_1}.
\]
Note that definitions of $\Gamma^A, \gamma^A, \Gamma^B, \gamma^B, H^A, H^B$ are the same when we only consider the subsystem as the total system.

Size-consistency within the RDM method requires:
\begin{eqnarray}
E_g (A\cdots B)& = & \min_{{\cal F}^{A\cdots B} \ni (\Gamma, \gamma)} \left \{ \sum_{ij} v^i_j \gamma^i_j + \sum_{i_1 i_2 j_1 j_2} w^{i_1 i_2}_{j_1 j_2} \Gamma^{i_1 i_2}_{j_1 j_2} \right \}, \nonumber \\ 
E_g (A)& = & \min_{{\cal F}^{A} \ni (\Gamma^A, \gamma^A)} \left \{ \sum_{ij} v^A{}^i_j \gamma^A{}^i_j + \sum_{i_1 i_2 j_1 j_2} w^A{}^{i_1 i_2}_{j_1 j_2} \Gamma^A{}^{i_1 i_2}_{j_1 j_2} \right \}, \nonumber \\ 
E_g (B)& = & \min_{{\cal F}^{B} \ni (\Gamma^B, \gamma^B)} \left \{ \sum_{ij} v^B{}^i_j \gamma^B{}^i_j + \sum_{i_1 i_2 j_1 j_2} w^B{}^{i_1 i_2}_{j_1 j_2} \Gamma^B{}^{i_1 i_2}_{j_1 j_2} \right \}, \nonumber 
\end{eqnarray}
and,
\begin{equation}
E_g(A \cdots B ) = E_g(A) + E_g(B). \label{sizecons}
\end{equation}
where ${\cal F}^{A\cdots B}, {\cal F}^{A}$ and ${\cal F}^{B}$ are sets of 1- and 2-RDMs satisfying certain $N$-representability conditions for the total system $A\cdots B$, for the subsystem $A$ and $B$, respectively.

Now, let us investigate the subsystems. To simplify the problem, and at the dissociation, each subsystem $A$ and $B$ should be the eigenstate of the number of electrons, thus we require each subsystem $A$ and $B$ to be the eigenstate of the number of electrons, i.e. $N_A$ and $N_B$ electrons in the subsystem $A$ and $B$, respectively,
\[
\sum_{i=1} {\gamma^A}^i_i = N_A, \quad \sum_{i,j=1}{\Gamma^A}^{ij}_{ij} = \frac{N_A(N_A-1)}{2}.
\]

A variational space for the subsystem ${\cal F}^A_{N_A}$ is defined by ${\cal F}^{A\cdots B}$ as follows:
{\small
\[
{\cal F}^A_{N_A} ({\cal F}^{A\cdots B})= \left \{(\Gamma^A, \gamma^A) \left |\begin{array}{l}  
\mbox{For $\Gamma \in {\cal F}^{A\cdots B}$,}\\ 
\mbox{ ${\Gamma^A}^{i_1 i_2}_{j_1 j_2} := \Gamma^{i_1 i_2}_{j_1 j_2}$ where $i_1, i_2, j_1, j_2$ correspond to  one-particle basis of $A$, }  \\
\mbox{${\gamma^A}^{i}_{j} := \gamma^{i}_{j} = \frac{2}{N-1} \sum_k \Gamma^{ik}_{jk}$ where $i, j$ correspond to one-particle basis of $A$, }\\
\mbox{ and $\sum_{i=1} {\gamma^A}^i_i = N_A, \quad \sum_{i,j=1}{\Gamma^A}^{ij}_{ij} = \frac{N_A(N_A-1)}{2}$.}
\end{array}
\right.
\right \}.
\]
}

In a similar manner, we can define ${\cal F}^B_{N_B}$. Finally, we consider the size-consistency problem eqn. (\ref{sizecons}) which is expressed as follows:
\begin{eqnarray}
& & \min_{{\cal  F}^{A\cdots B} \ni \Gamma, \gamma} \left \{ \sum_{ij} v^i_j \gamma^i_j + \sum_{i_1 i_2 j_1 j_2} w^{i_1 i_2}_{j_1 j_2} \Gamma^{i_1 i_2}_{j_1 j_2} \right \} = \nonumber \\ 
& & \min_{{\cal F}^{A}_{N_A}({\cal F}^{A\cdots B}) \ni (\Gamma^A, \gamma^A)} \left \{ \sum_{ij} v^A{}^i_j \gamma^A{}^i_j + \sum_{i_1 i_2 j_1 j_2} w^A{}^{i_1 i_2}_{j_1 j_2} \Gamma^A{}^{i_1 i_2}_{j_1 j_2} \right \} \nonumber \\
& & + \min_{{\cal F}^{B}_{N_B} ({\cal F}^{A\cdots B})\ni (\Gamma^B, \gamma^B)} \left \{ \sum_{ij} v^B{}^i_j \gamma^B{}^i_j + \sum_{i_1 i_2 j_1 j_2} w^B{}^{i_1 i_2}_{j_1 j_2} \Gamma^B{}^{i_1 i_2}_{j_1 j_2} \right \}  \label{eq}
\end{eqnarray}

Size-consistency disturbance happens when eq. (\ref{eq}) is violated. If only the necessary conditions are imposed; e.g. if we impose the $P$, $Q$, $G$ conditions as $N$-representability conditions,
the left hand side decreases \cite{Nakata09}:
\[
E_g(A\cdots B) \leq E_g(A) + E_g(B),
\]
and if only sufficient conditions are imposed; e.g. if we use ${\cal F}^{A\cdots B}$, ${\cal F}^A_{N_A}({\cal F}^{A\cdots B})$ and ${\cal F}^B_{N_B}({\cal F}^{A\cdots B})$ as a
set of 2-RDMs from the SDCI wavefunctions, the left hand side increases:
\[
E_g(A\cdots B) \geq E_g(A) + E_g(B).
\]

\subsection{The lemma and theorem for matrices}
In this subsection we show the lemma and the theorem of matrices.
\begin{lemm}
If $P$ is an $n\times n$ symmetric matrix and positive semidefinite, its $m\times m$ ($n > m$) principal submatrix $A_m$ is positive semidefinite.
\end{lemm}
\begin{pro}
Proof is given in \cite{MatAnalysis}.
\end{pro}

\begin{theo}
Let $A$ and $B$ be $n\times n$ Hermitian matrices. If 
$(U^\dag A U)_{ii}=(U^\dag B U)_{ii}, \ \forall i\in\{1,2,\ldots,n\}$ is
valid for all unitary matrices $U$, then $A=B$.
\end{theo}
\begin{pro}
Let us prove by induction on the dimension of the matrix.
If $n=1$, the result is trivial. Suppose it is valid for all
matrices of size $n-1$, consider all $n\times n$ unitary
matrix $U$ such that $U_{in}=U_{ni}=0 \ (1\leq i \leq n-1)$. From the
induction hypothesis, we conclude that $A_{ij}=B_{ij} \
(1\leq i \leq n-1, 1\leq j \leq n-1)$, and $A_{nn}=B_{nn}$.
In a similar manner, we can show that $A=B$ except for the
elements $(1,n)$ and $(n,1)$, if $U_{in}=U_{ni}=0 \
(2\leq i \leq n)$. Now, let us call the first row of an
arbitrary unitary matrix $U$ by $u_1$ and its last row 
by $u_n$. Then, $U^\dag (A-B)U=
(A_{1n}-B_{1n})u_1^\dag u_n + (A_{n1}-B_{n1})u_n^\dag u_1
=(A_{1n}-B_{1n})u_1^\dag u_n + ((A_{1n}-B_{1n})u_1^\dag u_n)^\dag$.
In other words, from our hypothesis, the diagonal elements of this
matrix should have zero real parts for any choice of unitary
matrix $U$. Therefore, $A_{1n}=B_{1n}=A_{n1}^\dag=B_{n1}^\dag$. $\blacksquare$
\end{pro}

\subsection{A set of approximate $N$-representable 2-RDMs satisfying size-consistency}

We further investigate the nature of ${\cal F}^A_{N_A}({\cal \tilde E}_N^{A\cdots B})$. From Lemma 1, $\Gamma^A \in {\cal F}^A_{N_A}({\cal \tilde E}_N^{A\cdots B}) $ is positive semidefinite, since $\Gamma$ is positive semidefinite and $\Gamma^A$ is the principal submatrix of $\Gamma$. Moreover, subsystem $Q^A$ and $G^A$ matrices defined as follows are also trivially positive semidefinite:
\[
{Q^A}^{i_1 i_2}_{j_1 j_2} = {Q}^{i_1 i_2}_{j_1 j_2} \mbox{ \hspace{1cm} where $i_1, i_2, j_1, j_2$ correspond to the one-particle basis of $A$,}
\]
and
\[
{G^A}^{i_1 i_2}_{j_1 j_2} = {G}^{i_1 i_2}_{j_1 j_2} \mbox{ \hspace{1cm} where $i_1, i_2, j_1, j_2$ correspond to the one-particle basis of $A$.}
\]

The RDM method for the subsystem $A$ then becomes:
\[
\min_{{\cal F}^{A}_{N_A}({\cal \tilde E}_N^{A\cdots B}) \ni (\Gamma^A, \gamma^A)} \left \{ \sum_{ij} v^A{}^i_j \gamma^A{}^i_j + \sum_{i_1 i_2 j_1 j_2} w^A{}^{i_1 i_2}_{j_1 j_2} \Gamma^A{}^{i_1 i_2}_{j_1 j_2} \right \}.
\]

Inconsistency arises since ${\cal F}^{A}_{N_A}({\cal \tilde E}_N^{A\cdots B})$ and ${\cal \tilde E}_{N_A}^{A}$ are different, and is numerically shown to that ${\cal F}^{A}_{N_A}({\cal \tilde E}_N^{A\cdots B}) \supset {\cal \tilde E}_{N_A}^{A}$ \cite{Nakata09}. Let us investigate $G$ condition in the subsystem $A$, 
\[
 {G^A}^{i_1 i_2}_{j_1 j_2} = \delta^{i_2}_{j_2} {\gamma^A}^{i_1}_{j_1} - 2 \Gamma^A{}^{i_1 j_2}_{j_1 i_2},
\]
where $\delta$ is the kronker's delta.
This $G^A$ matrix is apparently positive semidefinite. 
However, $\gamma^A$ carries some information from the total system $A\cdots B$. If we use only the variables in subsystem $A$, we need to use 1-RDM $\tilde \gamma^A$, taking the partial trace of $\Gamma^A$ of subsystem $A$, {\it not} $\Gamma$ of the 
total system $A\cdots B$:
\[
 {\tilde \gamma^A}{}^i_j =\frac{2}{N_A-1}\sum_{k=1}{\Gamma^A}^{ik}_{jk},
\]
then we consider the $G$-condition of the subsystem $A$ by defining $\tilde G^A$ matrix as follows:
\[
 {\tilde G^A}{}^{i_1 i_2}_{j_1 j_2} = \delta^{i_2}_{j_2} {\tilde \gamma^A}{}^{i_1}_{j_1} - 2 \Gamma^A{}^{i_1 j_2}_{j_1 i_2}.
\]
Semidefiniteness of this $\tilde G^A$ should serve as the $G$-condition in subsystem $A$. However, $\tilde \gamma^A$ may not be consistent with $\gamma^A$. This inconsistency causes size-inconsistency in the RDM method. The RDM method finds a lower energy value by breaking condition (d), consistency between 1- and 2-RDM, of subsystem $A$ (or $B$). Nevertheless, the following theorem establishes consistency between $\gamma^A$ and $\tilde \gamma^A$. 

\begin{theo}
If (possibly non-) $N$-representable 2-RDM $\Gamma$ of the whole system satisfies unitary invariant diagonal $N$-representability conditions, and subsystem $\Gamma^A$ is the eigenstate of the number of electrons, then, ${\tilde \gamma^A}{}^i_j := \frac{2}{N_A-1} \sum_{k=1}{\Gamma^A}^{ik}_{jk} = \gamma^A{}^{i}_{j}$ is valid.
\end{theo}

\begin{pro}
It is enough to show that ${\tilde \gamma^A}{}^i_i = \gamma^A{}^{i}_{i}$ for a specific choice on the one-particle basis. This is because the unitary invariant $N$-representability of $\Gamma$, and the expression of each linear inequality corresponding on the diagonal $N$-representability condition is independent of the choice of one-particle basis.
Then, ${\tilde \gamma^A}{}^i_j = \gamma^A{}^{i}_{j}$ is thus derived from Theorem 2.

Let us prove ${\tilde \gamma^A}{}^i_i = \gamma^A{}^{i}_{i}$. From the diagonal $N$-representability condition, there exists $|\Psi\rangle$ which reduces to the diagonal elements of $\Gamma$, 
\[
\Gamma^{ij}_{ij} = \frac{1}{2} \langle \Psi | a^\dagger_i a^\dagger_j a_j a_i | \Psi \rangle.
\]
We can choose such $|\Psi\rangle$ as 
\[
|\Psi \rangle = \sum_p w_p |\Phi_p\rangle,
\]
where $|\Phi_p\rangle$ are the Slater determinants, and $w_p$ is a non-negative value with unit sum ($\sum_p |w_p|^2 =1$) while $\phi_p$ is an arbitrary phase factor \cite{AyersDavidson07}.
 
Since we employ separate one-particle basis in $A$ and $B$, $|\Phi_p\rangle$ can be decomposed into $|\Phi^A_q \Phi^B_r \rangle := |\Phi^A_q \rangle \otimes |\Phi^B_r \rangle = |\Phi_p\rangle$, where $|\Phi^A_q\rangle$ and $|\Phi^B_r\rangle$ are the Slater determinants in each subsystem. In this way,
 the total wave function $|\Psi\rangle$ can be written as:
\begin{equation}
|\Psi \rangle = \sum_{pq} \sqrt{w_{pq}} e^{i\phi_{pq}}|\Phi_p^A \Phi_q^B\rangle,  \label{wfAB}
\end{equation}
where $w_{pq}$ is a non-negative value with unit sum ($\sum_{p,q} |w_{pq}|^2 =1$), and $\phi_{pq}$ is an arbitrary phase factor. Now, let us show that an ensemble $\sum_{pq} w_{pq} |\Phi^A_p\rangle \langle \Phi^A_p|$ that reproduces the diagonal elements of $\Gamma^A$.
\begin{eqnarray}
\Gamma^A{}^{ij}_{ij} & = & \frac{1}{2} \langle \Psi | a^\dagger_i a^\dagger_j  a_j a_i | \Psi \rangle \nonumber  \\
& = & \frac{1}{2} \sum_{pqrs} \sqrt{w_{pq}}e^{i\phi_{pq}} \sqrt{w_{rs}}e^{-i\phi_{rs}} \langle \Phi^A_r \Phi^B_s | a^\dagger_i a^\dagger_j  a_j a_i | \Phi^A_p \Phi^B_q \rangle \nonumber \\
& = & \frac{1}{2} \sum_{pqrs} \sqrt{w_{pq}}e^{i\phi_{pq}} \sqrt{w_{rs}}e^{-i\phi_{rs}} \langle \Phi^A_r | a^\dagger_i a^\dagger_j  a_j a_i | \Phi^A_p \rangle \langle \Phi^B_s  | \Phi^B_q\rangle \nonumber\\
& = &  \frac{1}{2} \sum_{pq} w_{pq} \langle \Phi^A_p | a^\dagger_i a^\dagger_j  a_j a_i | \Phi^A_p \rangle \nonumber
\end{eqnarray}
and
\begin{eqnarray}
{\rm tr} \left ( \frac{1}{2} a^\dagger_i a^\dagger_j a_j a_i \sum_{pq} w_{pq} |\Phi^A_p\rangle \langle \Phi^A_p| \right ) & = & \frac{1}{2} \sum_{pq} w_{pq} \langle \Phi^A_p | a^\dagger_i a^\dagger_j  a_j a_i | \Phi^A_p \rangle \nonumber\\
& = & \Gamma^A{}^{ij}_{ij}. \nonumber
\end{eqnarray}
$\tilde \gamma^A{}^{i}_{i}$ can be derived from $\sum_{pq} w_{pq} |\Phi^A_p\rangle \langle \Phi^A_p|$
\[
\tilde \gamma^A{}^{i}_{i} := \frac{2}{N_A-1} \sum_k \Gamma^A{}^{ik}_{ik} = 
{\rm tr} \left ( a^\dagger_i a_i \sum_{pq} w_{pq} |\Phi^A_p\rangle \langle \Phi^A_p|\right) = \sum_{pq} w_{pq} \langle \Phi^A_p | a^\dagger_i a_i | \Phi^A_p \rangle .
\]
Then, $\gamma^A{}^{i}_{i}=\tilde \gamma^A{}^{i}_{i}$ can be shown as follows:
\begin{eqnarray}
\gamma^A{}^{i}_{i} & = & \frac{2}{N -1} \sum_j \frac{1}{2}\langle \Psi | a^\dagger_i a^\dagger_j a_j a_i | \Psi \rangle \nonumber \\
                   & = & \frac{2}{N -1} \sum_{jpqrs} \frac{1}{2} \sqrt{w_{rs}}\sqrt{w_{pq}} e^{i\phi_{pq}} e^{-i\phi_{rs}}\langle \Phi_r^A \Phi_s^B|  a^\dagger_i a^\dagger_j a_j a_i |  \Phi_p^A \Phi_q^B\rangle \nonumber \\
                   & = & \frac{2}{N -1} \sum_{jpq} \frac{1}{2} w_{pq} \langle \Phi_p^A \Phi_q^B| a^\dagger_i a^\dagger_j a_j a_i |\Phi_p^A \Phi_q^B\rangle \nonumber \\
                   & = & \frac{2}{N -1} \left ( \sum_{j\in A, pq} \frac{1}{2} w_{pq} \langle \Phi_p^A \Phi_q^B| a^\dagger_i a^\dagger_j a_j a_i |\Phi_p^A \Phi_q^B\rangle  +  \sum_{j\in B, pq} \frac{1}{2} w_{pq}\langle \Phi_p^A \Phi_q^B| a^\dagger_i a^\dagger_j a_j a_i | \Phi_p^A \Phi_q^B\rangle \right ) \nonumber \\
                   & = & \frac{2}{N -1} \left ( \sum_{j\in A, pq} \frac{1}{2} w_{pq} \langle \Phi_p^A | a^\dagger_i a^\dagger_j a_j a_i |\Phi_p^A \rangle  +  \sum_{pq} \frac{N_B}{2} w_{pq} \langle \Phi_p^A | a^\dagger_i  a_i | \Phi_p^A \rangle \right ) \nonumber \\
                   & = & \frac{2}{N -1} \left ( \sum_j \Gamma^A{}^{ij}_{ij} + \frac{N_B}{2}\tilde \gamma^A{}^{i}_{i}\right ) \nonumber \\
                   & = & \frac{2}{N -1} \left ( \frac{N_A-1}{2}\tilde \gamma^A{}^{i}_{i} + \frac{N_B}{2}\tilde \gamma^A{}^{i}_{i} \right ) \nonumber \\
               & = & \tilde \gamma^A{}^{i}_{i}. \nonumber
\end{eqnarray}
$\blacksquare$
\end{pro}
A physical interpretation in Theorem 3 requires each subsystem energy to be evaluated by the variables or information of each subsystem.

Finally, we focus on Theorem 4.
\begin{theo}
The RDM method is size-consistent if the following conditions are satisfied; (i) the variational set of 2-RDM is unitary invariant diagonal $N$-representable, (ii) calculated subsystems are the eigenstate of the number of electrons, (iii) the 2-RDMs satisfy the arbitrary choice of positive semidefinite type $N$-representability conditions from conditions $P$, $Q$ and $G$.
\end{theo}

\begin{pro}
Considering two non-interacting subsystems $A$ and $B$ being treated
as a combined system $A\cdots B$, the total energy of the combined system is now become the
sum of the total energy of the subsystems; $E(A \cdots B) = E(A) + E(B)$. 
We define the approximate $N$-representable set of 1- and 2-RDMs of the total system ${\cal \bar E}_N^{A\cdots B}$ as follows:
\[
{\cal \bar E}_N^{A\cdots B} = \left \{ (\Gamma,\gamma) \left | \begin{array}{l}
\mbox{trivial conditions (a) to (d) with conditions $P$, $Q$, and $G$,} \\
\mbox{and the unitary invariant diagonal $N$-representability conditions.}
\end{array} \right.
\right \}
\]
Then, we project the total system in subsystem $A$ by:
\[
{\cal \bar E}^A_{N_A} = \left \{\Gamma^A,\gamma^A \left | \begin{array}{l}
\mbox{ For ${\cal \bar E}_N^{A\cdots B} \ni \Gamma, \gamma$,} \\
\mbox{ ${\Gamma^A}^{i_1 i_2}_{j_1 j_2} = \Gamma^{i_1 i_2}_{j_1 j_2}$ where $i_1, i_2, j_1, j_2$ corresponds to the one-particle basis of $A$,} \\
\mbox{ ${\gamma^A}^{i}_{j} = \gamma^{i}_{j}$ where $i, j$ corresponds to the one-particle basis of $A$,}\\
\mbox{$\sum_{ij} {\Gamma^A}^{ij}_{ij} = N_A (N_A-1)/2$, and $\sum_i {\gamma^A}^{i}_{i} = N_A$.} \\
\end{array} \right.
\right \}
\]
and the variational space can be defined by using only the variable in subsystem $A$ or denoted as ${\cal \bar {\bar E}}_A^{N_A}$ as follows:
\[
{\cal \bar {\bar E}}^{A}_{N_A} = \left \{\Gamma^A,\gamma^A \left | \begin{array}{l}
\mbox{trivial condition (a) to (d) with the $P$, $Q$, and $G$ conditions,} \\
\mbox{and the unitary invariant diagonal $N$-representability conditions.}
\end{array} \right.
\right \}
\]
It is sufficient to demonstrate that ${\cal \bar E}_A^{N_A} = {\cal \bar {\bar E}}_A^{N_A}$. First, we show that 
${\cal \bar E}_{N_A}^A \subseteq {\cal \bar {\bar E}}^{A}_{N_A}$. It is easy to see that (a) Hermite condition, (b) anti-symmetricity and (c) trace conditions are satisfied for either 1- or 2-RDMs of ${\cal \bar E}^A_{N_A}$. From Theorem 3, partial trace condition holds for the subsystems, thus validating (d) as well. Now let us construct the $Q^A$ and $G^A$ matrices as follows:
\[
\mbox{ ${Q^A}^{i_1 i_2}_{j_1 j_2} = Q^{i_1 i_2}_{j_1 j_2}$ where $i_1, i_2, j_1, j_2$ correspond to the one-particle basis of $A$,}
\]
and 
\[
\mbox{ ${G^A}^{i_1 i_2}_{j_1 j_2} = G^{i_1 i_2}_{j_1 j_2}$ where $i_1, i_2, j_1, j_2$ correspond to the one-particle basis of $A$.}
\]
Matrices $P^A$, $Q^A$ and $G^A$ are all positive semidefinite, because from Lemma 1, all principal submatrices are positive semidefinite. Again, from Theorem 3, $\gamma^A$ is derived from $\Gamma^A$, thus matrices $Q^A$ and $G^A$ of subsystem $A$ are expressed via $\Gamma^A$ and $\gamma^A$. Furthermore, both sets satisfy the diagonal $N$-representability conditions to yield ${\cal \bar E}^A_{N_A} \subseteq {\cal \bar {\bar E}}^A_{N_A}$.
Next, ${\cal \bar {\bar E}}^A_{N_A} \subseteq {\cal \bar E}^A_{N_A}$, where we construct the 1 and 2-RDM $\gamma$ and $\Gamma$ of the total system from subsystem 1- and 2-RDM of $A$ and $B$; $\gamma^A$, $\gamma^B$, $\Gamma^A$ and $\Gamma^B$, respectively by:
\[
\gamma = \left \{ \begin{array}{l}
\mbox{ ${\gamma}^{i}_{j} = \gamma^A{}^{i}_{j}$ where $i, j$ correspond to the one-particle basis of $A$,}\\
\mbox{ ${\gamma}^{i}_{j} = \gamma^B{}^{i}_{j}$ where $i, j$ correspond to the one-particle basis of $B$,}\\
\mbox{ ${\gamma}^{i}_{j} = 0$ where otherwise.}
\end{array} \right.
\]
and
\[
\Gamma = \left \{ \begin{array}{ll}
\mbox{ ${\Gamma}^{i_1 i_2}_{j_1 j_2} = \Gamma^A{}^{i_1 i_2}_{j_1 j_2} $ } & \mbox{where $i_1, i_2, j_1, j_2$ correspond to the one-particle basis of $A$} \\
\mbox{ ${\Gamma}^{i_1 i_2}_{j_1 j_2} = \Gamma^B{}^{i_1 i_2}_{j_1 j_2} $ } & \mbox{where $i_1, i_2, j_1, j_2$ correspond to the one-particle basis of $B$} \\
\mbox{ ${\Gamma}^{i_1 i_2}_{j_1 j_2} = \frac{1}{2}\gamma^A{}^{i_1}_{j_1} \gamma^B{}^{i_2}_{j_2}$} & \mbox{where $i_1, j_1$ correspond to the one-particle basis of $A$,}\\
& \mbox{and $i_2, j_2$ corresponds to the one-particle basis of $B$} \\
\mbox{ ${\Gamma}^{i_1 i_2}_{j_1 j_2} = \frac{1}{2}\gamma^A{}^{i_2}_{j_2} \gamma^B{}^{i_1}_{j_1} $} & \mbox{where $i_2, j_2$ correspond to the one-particle basis of $A$,}\\
& \mbox{and $i_1, j_1$ corresponds to the one-particle basis of $A$} \\
\mbox{ ${\Gamma}^{i_1 i_2}_{j_1 j_2} = - \frac{1}{2}\gamma^A{}^{i_1}_{j_2} \gamma^B{}^{i_2}_{j_1}$} & \mbox{where $i_1, j_2$ correspond to the one-particle basis of $A$,}\\
& \mbox{and $i_2, j_1$ correspond to the one-particle basis of $B$} \\
\mbox{ ${\Gamma}^{i_1 i_2}_{j_1 j_2} = - \frac{1}{2}\gamma^A{}^{i_2}_{j_1} \gamma^B{}^{i_1}_{j_2}$} & \mbox{where $i_2, j_1$ correspond to the one-particle basis of $A$,}\\
& \mbox{and $i_1, j_2$ correspond to the one-particle basis of $B$}\\
\mbox{${\Gamma}^{i_1 i_2}_{j_1 j_2} = 0$} & \mbox{otherwise}.
\end{array} \right.
\]
It is easy to show that $\Gamma$ and $\gamma$ satisfy (a), (b), (c) and (d). The 2-RDM $\Gamma$ has a structure containing
three blocks; (i) where indices correspond to subsystem $A$, (ii) indices correspond to subsystem $B$, (iii) indices correspond to both subsystems $A$ and $B$. For condition $P$, it is sufficient to express these three matrices as positive semidefinite. Apparently,
the first two blocks are positive semidefinite by their construction. The third block is always $N$-representable from the unitary invariant diagonal $N$-representability conditions (an explicit representation of the third block is derived from eq. (\ref{wfAB}) and Theorem 2 employing an one-particle basis which diagonalizes subsystems $A$ and $B$), thus $\Gamma$ satisfies condition $P$. We can also show the positive semidefiniteness of matrices $Q$ and $G$ of the total system as well. Thus $(\Gamma, \gamma) \in {\cal \bar E}_N^{A\cdots B}$, then ${\cal \bar {\bar E}}^A_{N_A} \subseteq {\cal \bar E}^A_{N_A}$ holds. Combining these two results,
we obtain ${\cal \bar {\bar E}}^A_{N_A} = {\cal \bar E}^A_{N_A}$. Therefore, the total energy becomes the sum of each subsystem. $\blacksquare$
\end{pro}

Here we have two corollaries.
\begin{coll}
Size-consistency is also satisfied by the RDM method employing 2-RDMs as long as the unitary invariant diagonal $N$-representability conditions with $P$, $Q$, $G$, $T1$ and $T2^\prime$ condition satisfied, assuming that each system is an eigenstate of the number of electrons at the minimum.
\end{coll}

\begin{coll}
Size-consistency is satisfied by the RDM method employing 2-RDMs as long as the unitary invariant diagonal $N$-representability conditions with an arbitrary combination of conditions $P$, $Q$, $G$, $T1$, and $T2^\prime$ are satisfied, assuming that each system is the eigenstate of the number of electrons at the minimum.
\end{coll}



\section{Discussion and conclusion}
We have showed size-consistency can be reestablished by the RDM method using unitary invariant diagonal $N$-representability conditions in 2-RDM. This is an extension of the diagonal $N$-representability conditions.

We also assumed that in the RDM method, electrons are distributed to each subsystem appropriately. This requirement looks stringent, as Van Aggelen {\it et al.} have showed that an incorrect distribution exists at the dissociation limit. However, we concur on the finding that the unitary invariant diagonal $N$-representability conditions results in the correct distribution of electrons at the non-interacting limit. Thus, we can omit this assumption. Note that we usually employ ensemble $N$-representability. In such a situation, even if we force to employ the complete ensemble to satisfy $N$-representability, each subsystem would become a fractional number of electrons. For example, consider the dissociation limit of a diatomic isonuclear system with an odd number of electrons, e.g., $\rm N_2^+$. It is not clear which $\rm N$ atom has seven electrons, and ends up as an ensemble of six to seven electrons. In this case, we need a pure $N$-representability condition, or we need to apply certain perturbation to break the degeneracy.

In practice, inclusion of the unitary invariant diagonal $N$-representability conditions is not a simple task. A possible way is finding a positive semidefinite type $N$-representability conditions corresponding to implying inequalities from the diagonal $N$-representability conditions. For example, condition $P$ implies $\Gamma^{ij}_{ij} \geq 0$ and it includes $\Gamma^{ij}_{ij}\geq 0 $ for all one-particle basis representations. Apparently condition $P$ is stronger than $\Gamma^{ij}_{ij} \geq 0$ since semidifiniteness of $P$-matrix include unitary rotation of two-particle operators. 
Likewise, condition $Q$ implies $1 - \gamma^i_i - \gamma^j_j  + \Gamma^{ij}_{ij} \geq 0$ and condition $G$-condition $\gamma^{i}_{i} - \Gamma^{ij}_{ij} \geq 0 $. Using anti-commutation relation of creation and annihilation operators 
\[
a^\dagger_i a_j  + a_j a^\dagger_i = \delta^{i}_{j},\mbox{\hspace{0.5cm}}
 a_i a_j  + a_j a_i = 0, \mbox{\,\,\, and \,\,\,} a^\dagger_i a^\dagger_j  + a^\dagger_j a^\dagger_i = 0, 
\]
where $\delta$ is the Kronecker's delta, we can verify these relation by taking the diagonal elements of $Q$ and $G$-matrices. For $Q$-matrix as follows:
\begin{eqnarray}
Q^{i_1 i_2}_{j_1 j_2} & = &
\langle \Psi | a_{i_1} a_{i_2} a^\dagger_{j_2} a^\dagger_{j_1} | \Psi \rangle \nonumber \\
& = &  (\delta^{i_1}_{j_1} \delta^{i_2}_{j_2} 
 - \delta^{i_1}_{j_2} \delta^{i_2}_{j_1}) 
- (\delta^{i_1}_{j_1} \gamma^{i_2}_{j_2} + 
  \delta^{i_2}_{j_2} \gamma^{i_1}_{j_1} )
+ (\delta^{i_1}_{j_2} \gamma^{i_2}_{j_1} + 
   \delta^{i_2}_{j_1} \gamma^{i_1}_{j_2})
   - 2 \Gamma^{i_1 i_2}_{j_1 j_2}, \nonumber
\end{eqnarray}
and non-negativity of diagonal elements of $Q$-matrix implies following inequality,
\begin{eqnarray}
Q^{ij}_{ij} & = &  1 - \gamma^j_j - \gamma^i_i - 2\Gamma^{ij}_{ij} \geq 0. \nonumber
\end{eqnarray}
For $G$-matrix, we can show as follows:
\begin{eqnarray}
G^{i_1 i_2}_{j_1 j_2} & = & \langle \Psi | a^\dagger_{i_1} a_{i_2} a^\dagger_{j_2} a_{j_1} | \Psi \rangle \nonumber \\
 & = &  (\delta^{i_2}_{j_2} \gamma^{i_1}_{j_1} - 2 \Gamma^{i_1 j_2}_{j_1 i_2}) \geq 0, \nonumber
\end{eqnarray}
and non-negativity of diagonal elements of $G$-matrix implies following inequality,
\begin{eqnarray}
G^{ij}_{ij} & = &  \gamma^i_i - 2\Gamma^{ij}_{ij} \geq 0 \nonumber.
\end{eqnarray}
Likewise, for $T1$-matrix, we can show as follows:
\begin{eqnarray}
(T1)^{i_1 i_2 i_3}_{j_1 j_2 j_3} & = & \langle \Psi |a^\dagger_{i_1}a^\dagger_{i_2}a^\dagger_{i_3} a_{j_3}a_{j_2}a_{j_1}+ a_{j_1}a_{j_2}a_{j_3} a^\dagger_{i_3}a^\dagger_{i_2}a^\dagger_{i_1}  |\Psi \rangle\nonumber \\
 & =  & 2 ( \delta^{j_2}_{i_2} \Gamma^{i_1 i_3}_{j_1 j_3} +
\delta^{j_3}_{i_3} \Gamma^{i_2 i_1}_{j_2 j_1} +
\delta^{j_1}_{i_1} \Gamma^{i_3 i_2}_{j_3 j_2} +
\delta^{j_1}_{i_3} \Gamma^{i_2 i_1}_{j_3 j_2} \nonumber \\ 
 & &-\delta^{j_2}_{i_3} \Gamma^{i_2 i_1}_{j_3 j_1} -
\delta^{j_3}_{i_2} \Gamma^{i_3 i_1}_{j_1 j_2} +
\delta^{j_3}_{i_1} \Gamma^{i_3 i_2}_{j_2 j_1} -
\delta^{j_1}_{i_2} \Gamma^{i_3 i_1}_{j_3 j_2} -
\delta^{j_2}_{i_1} \Gamma^{i_3 i_2}_{j_3 j_1} ) \nonumber \\
 &&
+\delta^{j_3}_{i_1} \delta^{j_1}_{i_3} \gamma^{i_2}_{j_2} -
 \delta^{j_2}_{i_1} \delta^{j_1}_{i_3} \gamma^{i_2}_{j_3} -
\delta^{j_2}_{i_1} \delta^{j_3}_{i_2} \gamma^{i_3}_{j_1} +
\delta^{j_1}_{i_1} \delta^{j_3}_{i_2} \gamma^{i_3}_{j_2} +
\delta^{j_3}_{i_1} \delta^{j_2}_{i_2} \gamma^{i_3}_{j_1} \nonumber \\ 
 &&-\delta^{j_1}_{i_1} \delta^{j_2}_{i_2} \gamma^{i_3}_{j_3} +
\delta^{j_2}_{i_2} \delta^{j_1}_{i_3} \gamma^{i_1}_{j_3} -
\delta^{j_2}_{i_2} \delta^{j_3}_{i_3} \gamma^{i_1}_{j_1} +
\delta^{j_1}_{i_2} \delta^{j_3}_{i_3} \gamma^{i_1}_{j_2} +
\delta^{j_2}_{i_1} \delta^{j_3}_{i_3} \gamma^{i_2}_{j_1} \nonumber \\ 
 &&-\delta^{j_1}_{i_1} \delta^{j_3}_{i_3} \gamma^{i_2}_{j_2} +
\delta^{j_3}_{i_2} \delta^{j_2}_{i_3} \gamma^{i_1}_{j_1} -
\delta^{j_1}_{i_2} \delta^{j_2}_{i_3} \gamma^{i_1}_{j_3} -
\delta^{j_3}_{i_1} \delta^{j_2}_{i_3} \gamma^{i_2}_{j_1} +
\delta^{j_1}_{i_1} \delta^{j_2}_{i_3} \gamma^{i_2}_{j_3} \nonumber \\ 
 &&-\delta^{j_3}_{i_2} \delta^{j_1}_{i_3} \gamma^{i_1}_{j_2} -
\delta^{j_3}_{i_1} \delta^{j_1}_{i_2} \gamma^{i_3}_{j_2} +
\delta^{j_2}_{i_1} \delta^{j_1}_{i_2} \gamma^{i_3}_{j_3} -
\delta^{j_2}_{i_1} \delta^{j_1}_{i_2} \delta^{j_3}_{i_3} -\delta^{j_3}_{i_1} \delta^{j_2}_{i_2} \delta^{j_1}_{i_3} \nonumber \\ 
 &&+\delta^{j_2}_{i_1} \delta^{j_3}_{i_2} \delta^{j_1}_{i_3} +\delta^{j_3}_{i_1} \delta^{j_1}_{i_2} \delta^{j_2}_{i_3} +\delta^{j_1}_{i_1} \delta^{j_2}_{i_2} \delta^{j_3}_{i_3} -\delta^{j_1}_{i_1} \delta^{j_3}_{i_2} \delta^{j_2}_{i_3}. \nonumber 
\end{eqnarray}
Then, the non-negativity of the diagonal part of $T1$-matrix
\begin{eqnarray}
(T1)^{i j k}_{i j k} & = & 2 \Gamma^{i k}_{i k} + 2 \Gamma^{j i}_{j i} + 2 \Gamma^{k j}_{k j} - \gamma^{k}_{k} - \gamma^{i}_{i} - \gamma^{j}_{j}  + 1 \geq 0\nonumber
\end{eqnarray}
implies Condition VI: $1 - \gamma^{i}_{i} - \gamma^{j}_{j} - \gamma^{k}_{k} 
+ 2 \Gamma^{ij}_{ij} + 2 \Gamma^{ik}_{ik}  + 2 \Gamma^{jk}_{jk}$.
For $T2$-matrix,
\begin{eqnarray}
(T2)^{i_1 i_2 i_3}_{j_1 j_2 j_3} & = & \langle \Psi |a^\dagger_{i_1}a^\dagger_{i_2}a_{i_3} a^\dagger_{j_3}a_{j_2}a_{j_1}+ a^\dagger_{j_3}a_{j_2}a_{j_1} a^\dagger_{i_1}a^\dagger_{i_2}a_{i_3} |\Psi \rangle\nonumber \\
& = & 2 \delta^{i_3}_{j_3} \Gamma^{i_1 i_2}_{j_1 j_2} - \delta^{j_1}_{i_1} \Gamma^{j_3 i_2}_{i_3 j_2} +\delta^{j_2}_{i_1}
\Gamma^{j_3 i_2}_{i_3 j_1} +\delta^{j_1}_{i_2} \Gamma^{j_3 i_1}_{i_3 j_2} -\delta^{j_2}_{i_2} 
\Gamma^{j_3 i_1}_{i_3 j_1}) \nonumber \\ 
 &&+\delta^{j_2}_{i_2} \delta^{j_1}_{i_1} \gamma^{j_3}_{i_3} -\delta^{j_1}_{i_2} \delta^{j_2}_{i_1} \gamma^{j_3}_{i_3}, \nonumber
\end{eqnarray}
and the diagonal elements of $T2$-matrix is following:
\begin{eqnarray}
(T2)^{ijk}_{ijk} & = & 2 \Gamma^{ij}_{ij} - 2\Gamma^{kj}_{kj} 
 -2 \Gamma^{k i}_{k i} + \gamma^{k}_{k} \geq 0, \nonumber
\end{eqnarray}
and its non-negativity implies Condition VII: $\gamma^{i}_{i} - 2 \Gamma^{ij}_{ij} - 2 \Gamma^{ik}_{ik}  + 2 \Gamma^{jk}_{jk}\geq 0$, by just reordering the indices, of Weinhold-Wilson inequality, respectively. Non-negativity of $T2^\prime$ matrix implies Condition VII and $\gamma^{i}_i \geq 0$ \cite{DIAGONALREP}. There exists such a correspondence for each condition and inequality \cite{AyersDavidson07}. By just adding all positive semidefinite type $N$-representability conditions (e.g. $P$, $Q$, $G$, $T1$ and $T2^\prime$ etc.) is enough to satisfy the unitary invariant diagonal $N$-representability conditions. 

Interestingly, the unitary invariant diagonal $N$-representability has already been proposed to be not only for a necessity but also for a sufficiency by Weinhold \cite{Weinhold68}. In fact, this has recently been proven at the 50th Sanibel symposium by Weiner \cite{Weiner2010}. However, as far as the author's knowledge, a complete proof has never been in the literature. Of course if it is ture, size-consistency is automatically satisfied. The importance of our results is a direct proof using unitary invariant $N$-representability conditions. Assuming the unitary invariant diagonal $N$-representability conditions are necessary and sufficient, we can express the following corollary:
\begin{coll}
The complete set of positive semidefinite type $N$-representability conditions analogous to the diagonal representability conditions will be a complete set of the $N$-representability conditions.
\end{coll}
Such an extension is preferred because we need only the information from 2-RDM for the RDM method and not higher order conditions \cite{ERDAHLJIN00}, we can still formulate the problem by semidefinite programming, and the approximate $N$-representable set equates asymptotically exactly by adding $N$-representability conditions from the diagonal $N$-representability conditions.
A positive semidefinite type condition seems to be much more stronger than the corresponding linear inequality \cite{Nakata04,Zhao04,DIAGONALREP}. Thus we expect rapid convergence. 

Finding practical methods for incorporating the diagonal $N$-representability conditions or the unitary invariant diagonal $N$-representability conditions is a challenge that warrants future investigation.

\begin{acknowledgments}
M.~N. was supported by the Special Postdoctoral 
Researchers' Program of RIKEN, and the study is partially supported by Grant-in-Aid for Scientific Research (B) 21300017. M.~N. is very thankful to Dr. Koji Yasuda at Nagoya University, Dr. Mituhiro Fukuda at Tokyo Tech, and Dr. Bastiaan Braams at the International Atomic Energy Agency for critical and constructive comments in realizing this study.
\end{acknowledgments}

\end{document}